\documentclass[12pt]{article}
\usepackage{amsfonts}
\usepackage[centertags]{amsmath}
\usepackage{amssymb}

\def\ba{\begin{array}}
\def\ea{\end{array}}
\def\be{\begin{equation}}
\def\ee{\end{equation}}
\def\bea{\begin{eqnarray}}
\def\eea{\end{eqnarray}}
\usepackage[verbose,colorlinks=true,naturalnames=true,linkcolor=blue,]{hyperref}
\newcounter{rown}

\begin{document}

\title{$N={\frac{1}{2}} $ Deformations of Chiral Superspaces from\\
New Quantum Poincar\'{e} and Euclidean Superalgebras}
\author{A. Borowiec$^{1}$, J. Lukierski$^{1}$, M. Mozrzymas$^{1}$ and V.N.
Tolstoy$^{1,2}$ \\
%EndAName
\\
$^{1}$Institute for Theoretical Physics, \\
University of Wroc{\l }aw, pl. Maxa Borna 9, \\
50--205 Wroc{\l }aw, Poland\\
\\
$^{2}$Lomonosov Moscow State University,\\
Skobeltsyn Institute of Nuclear Physics, \\
Moscow 119991, Russian Federation}
\date{}
\maketitle

\begin{abstract}
We present a large class of supersymmetric classical $r$-matrices, describing the
supertwist deformations of Poincar\'{e} and Euclidean superalgebras. We consider in
detail new family of four supertwists of $N=1$ Poincar\'{e} superalgebra and provide as
well their Euclidean counterpart. The proposed supertwists are better adjusted to the
description of deformed $D=4$ Euclidean supersymmetries with independent left-chiral and
right-chiral supercharges. They lead to new quantum superspaces, obtained by the
superextension of twist deformations of spacetime providing Lie-algebraic
noncommutativity of space-time coordinates. In the Hopf-algebraic Euclidean SUSY
framework the considered supertwist deformations provide an alternative to the
$N={\frac{1}{2}}$ SUSY Seiberg's star product deformation scheme.
\end{abstract}

\setcounter{equation}{0}

\section{Introduction}
Basic theories of fundamental interactions (string theory, $M$-theory) are supersymmetric
and the framework of quantum deformations for relativistic systems should be
supersymmetrically extended. In this paper we shall describe the deformations of $D=4$ Poincare and Euclidean
supersymmetries and provide new models of deformed chiral
superspace, realized by twist quantization procedure.

There were considered two ways of embedding the noncommutative (super)space algebras into
the (super)symmetry framework, which were further used for the formulation of deformed
dynamical theories:

{\bf(a)} One postulates as a basic notion the noncommutative structures of deformed
(super)space coordinates, and subsequently one defines corresponding star products
representing the multiplication of deformed (super)fields. In such a scheme one keeps
unchanged the standard (super)Poincar\'{e} symmetries, and noncommutativity is
interpreted as introducing the breaking of standard relativistic (super)symmetries (see
e.g. \cite{DFR,Filk} for non-SUSY and \cite{Seib,Ivan,Ferrara,Buch1} for SUSY case). In
such a framework the simplest $N={1\over2}$ supersymmetric deformation was proposed by
Seiberg \cite{Seib}.

{\bf(b)} One introduces the star product describing the noncommutative structure of the
(super)space as derived from quantum-deformed Poincar\'{e}--Hopf (super)algebra. If the
quantum symmetry is generated by a twist factor, it provides explicit definition of the
star multiplication \cite{Oeckl}--\cite{Dimi}. In such framework the primary notion
defining the choice of deformation is given by quantum Poincar\'{e}--Hopf (super)algebra.
In particular in the case of triangular deformation, only the coalgebra is modified by
the twist factor $F$ via a similarity transformation\footnote{If we permit the twists
satisfying the cocycle condition modified by nontrivial co-associator $\Phi$, it was
argued by Drinfeld (\cite{Dr}; see also \cite{YZ}) that any quantum deformation of
Poincar\'{e} (super)algebra can be represented by (super)twist in the framework of
quasi-Hopf algebras. Here we shall consider standard framework of quantum groups with
twists satisfying standard two-cocycle condition.}. In such formulation the algebraic
relations describing noncommutativity structure of Minkowski (super)space are by
construction covariant under the transformations of quantum (super)Poincar\'{e} group. We
stress that if the deformation of standard relativistic (super)symmetry is obtained by
(super)twist factor then the whole deformation is located in the coalgebra sector and the
classical Lie (super) algebras describing spacetime (super)symmetry are not modified.

In this paper we shall employ the second approach with primary notion of quantum
Hopf-algebraic  (super)symmetries. For chosen supertwists which should be generated from
classical supersymmetric $r$-matrices  we derive star-multiplication rules as well the
noncommutativity relations for (super)space coordinates. Firstly we shall study the
classification of superextensions of the classical $r$-matrices respectively for the
Poincar\'{e} Lie algebra and its Euclidean counterpart. It appears that due to different
reality conditions for Lorentz and $\mathfrak{o}(4)$ spinors  the superextensions are
consistent either for Minkowski or for Euclidean metric.

Let us recall firstly the twist deformation of relativistic symmetries. If we provide a
twist function $F\in U(\mathcal{P}(3,1))\otimes U(\mathcal{P}(3,1))$ (where
$\mathcal{P}(3,1)$ is Poincar\'{e} Lie algebra) both the deformed Poincar\'{e} symmetries
and quantum deformations of spacetime coordinates are uniquely determined. The basic
example of a twisted Poincar\'{e} deformation is provided by the canonical (Moyal--Weyl)
twist \cite{{Oeckl}}--\cite{Wess} which preserves the constant values of the commutator
of noncommutative Minkowski coordinates (for examples of other Poincar\'{e} twists
providing more general covariant noncommutative space-times see e.g. \cite{LW}). We add
that twisting of Poincar\'{e} symmetries was used for obtaining the quantum covariant
formulation of noncommutative field theories (see e.g. \cite{Oeckl}, \cite{Bloch},
\cite{FW}--\cite{Asc}) as well as of noncommutative gravity (see e.g. \cite{ADMSW}).

Analogously, in supersymmetric relativistic theories the supertwist function $F$ with
values in graded tensor product $U(\mathcal{P}(3,1|1))\otimes U(\mathcal{P}(3,1|1))$
(where $\mathcal{P}(3,1|1)$ is the $N=1$ Poincar\'{e} Lie superalgebra) defines the
deformed Poincar\'{e}--Hopf supersymmetries as well as covariant quantum deformation of
superspace. The technique of twisted Poincar\'{e} supersymmetry extending to SUSY
theories the results of \cite{CKNT,Wess} has been already studied for Minkowski (see e.g.
\cite{KS,IKS}) as well as for Euclidean (see e.g. \cite{IS}--\cite{GW1203}) %%
supersymmetry\footnote{We observe that in Minkowski case there were often used in field
theoretic applications nonstandard twists which are not spanned by the generators of
Poincar\'{e} superalgebra (see e.g. \cite{Bal}--\cite{BP}). In particular there was
employed a twist factor defined as function of the odd covariant derivatives in the
superspace (see e.g. \cite{DRW,DR}) which do anticommute with supercharges and extend the
basis of Poincar\'{e} superalgebra by graded Abelian algebra. We shall restrict in this
paper to standard supertwists of Drinfeld type, depending only on the Poincar\'{e}
superalgebra generators.}.

Following the classification of $D=4$ Poincar\'{e} deformations by the classical
$r$-matrices \cite{Z97} we shall study their Euclidean counterpart and further the
supersymmetric extensions of Poincar\'{e} and Euclidean cases. The aim and novelty of our
approach is: \\
{\bf(i)} to show that in view of the known list for classical Poincar\'{e} $r$-matrices
presented in \cite{Z97} one can find corresponding list of $D=4$ Euclidean $r$-matrices,\\
{\bf(ii)} to consider the consistency of superextensions of classical $\mathfrak{o}(3,1)$
and $\mathfrak{o}(4)$ $r$-matrices with reality conditions defining respectively the
Poincar\'{e} and Euclidean superalgebras, \\
{\bf(iii)} to supersymmetrize twist deformations which provide the Lie-algebraic
deformations of the space-time coordinates, \\
{\bf(iv)} to demonstrate that in the framework of twisted Euclidean superalgebras one can
obtain also $N={\frac{1}{2}}$ superspace deformation proposed by Seiberg \cite{Seib}
however with the deformed superspace algebra covariant under quantum Euclidean
supersymmetries.

The plan of the paper is the following. In Section 2 we shall consider $N=1$ Poincar\'{e}
and Euclidean superalgebras and describe Minkowski and Euclidean reality structures based
respectively on conjugation and pseudoconjugation in a fermionic sector. In Section 3 we
firstly introduce Euclidean counterpart of $D=4$ Poincar\'{e} classical $r$-matrices and
then we classify their corresponding supersymmetric Poincar\'{e} and Euclidean classical
$r$-matrices. Further, in Section 4 we recall the standard Moyal--Weyl twist deformation
of a space-time and present new four Euclidean supertwists which provide in bosonic
sector the Lie-algebraic deformations of the spacetime. In Section 5 we shall present in
Euclidean case the corresponding $N={\frac{1}{2}}$ SUSY deformation of Euclidean chiral
superspace. We get the first examples of deformations of Euclidean superspace coordinates
containing the Lie-algebraic deformation in its even space-time sector. In our case due
to twist deformation only the half of odd coalgebra relations are deformed and in
alternative deformation scheme of Seiberg \cite{Seib} only the algebraic sector is
changed, with modified anticommutativity of antichiral supercharges. In Section 6 we
shall present conclusion and outlook.

\setcounter{equation}{0}
\section{$D=4$ Poincar\'{e} and Euclidean superalgebras}
Real Poincar\'{e} (Euclidean) Lie algebra $\mathcal{P}(3,1))=\mathfrak{o}(3,1)\ltimes
\mathbf{P}$ ($\mathcal{E}(4)=\mathfrak{o}(4)\ltimes\mathbf{P}$) is generated by the
Poincar\'{e} (Euclidean) fourmomenta\footnote{In order to shorten our presentation we
denote in the same way the tensorial indexes of Poincar\'{e} and Euclidean generators. In
conventional notation our  Euclidean tensorial index "0" is denoted as the index "4".}
$P_{\mu}\in\mathbf{P}$ ($\mu=0,1,2,3$), and the six Lorentz (Euclidean) rotations
$L_{\mu\nu}\in\mathfrak{o}(3,1)$ ($L_{\mu\nu}\in\mathfrak{o}(4)$) ($\mu,\nu=0,1,2,3$)
satisfying the standard relations:
\begin{eqnarray}\label{se1}
\begin{array}{rcccl}
[L_{\mu\nu},\,L_{\lambda\rho}]\!\!& =\!\! & i\bigl(g_{\nu\lambda}\,L_{\mu\rho}-
g_{\nu\rho}\,L_{\mu\lambda}+g_{\mu\rho}\,L_{\nu\lambda}-g_{\mu\lambda}\,
L_{\nu\rho}\bigr)~,\qquad L_{\mu\nu}\!\!& =\!\!& -L_{\nu\mu}~,%%
\\[12pt]
[L_{\mu\nu},\,P_{\rho}]\!\!&=\!\!&i\bigl(g_{\nu\rho}\,P_{\mu}-g_{\mu\rho}\,
P_{\nu}\bigr)~,\qquad\qquad\qquad\qquad\quad\;[P_{\mu},\,P_{\nu}]\!\!&=\!\!&0~,
\end{array}%
\end{eqnarray}
where the metric $g_{\mu\nu}$ is given by $(g_{\mu\nu})=(g_{\mu\nu}^{\mathcal{P}})=
\mathop{\rm diag}\,(1,-1,-1,-1)$ for the Poincar\'{e} case and ($(g_{\mu\nu})=
(g_{\mu\nu}^{\mathcal{E}})=\mathop{\rm diag}\,(-1,-1,-1,-1)$ for the Euclidean case) and
the reality conditions imposed:
\begin{eqnarray}\label{se2}
L_{\mu\nu}^{*}\!\!&=\!\!&L_{\mu\nu}~,\qquad\quad P^{*}_{\mu}\,=\,P_{\mu}~.%%
\end{eqnarray}
Here the ${^*}$-antiinvolution is represented by Hermitian conjugation\footnote{Physical
Poincar\'{e} $\mathcal{P}(3,1)$ and Euclidean $\mathcal{E}(4)$ Lie algebras can be
defined as two real forms of complex inhomogeneous algebra
$IO(4;\mathbb{C})=O(4;\mathbb{C})\ltimes\mathbf{P}_\mathbb{C}$ where
$\mathbf{P}_\mathbb{C}$ denotes complex four-translations.}.

\noindent $N=1$ Poincar\'{e} (Euclidean) superalgebra $\mathcal{P}(3,1|1)$
($\mathcal{E}(4|1)$) is generated by the Poincar\'{e} (Euclidean) algebra
$\mathcal{P}(3,1)$ ($\mathcal{E}(4)$) and the four complex spinor generators  $Q_\alpha$
and $\bar{Q}_{\dot{\alpha}}$ ($\alpha=1,2;\;\dot{\alpha}=\dot{1},\dot{2}$) satisfying the
following relations\footnote{Again in order to have a compact presentation we did adjust
the notation in Euclidean superalgebra case to the standard formulation of $N=1$
Poincar\'{e} superalgebra. In Euclidean case due to the spinorial covering
$\overline{O(4)}=SU_{L}(2)\otimes SU_{R}(2)$ the supercharges $Q_{\alpha}$ and
$\bar{Q}_{\dot{\alpha}}$ are two independent $SU_{L}(2)$ and $SU_{L}(2)$ spinors. In a
transparent notation the Euclidean supercharges $(Q_{\alpha},\;\bar{Q}_{\dot{\alpha}})$
can be denoted as in \cite{LN84} by $(Q_{\alpha;},\;\bar{Q}_{;\dot\alpha}=Q^{;\alpha})$.}: %
\begin{eqnarray}\label{se3}
\begin{array}{rcccl}
\{Q_{\alpha},\,Q_{\beta}\}\!\!&=\!\!&\{\bar{Q}_{\dot{\alpha}},\,
\bar{Q}_{\dot{\beta}}\}\,= \,0,\qquad\{Q_{\alpha},\,\bar{Q}_{\dot{\beta}}\}\!\!&=\!\!&
2(\sigma^{\mu})_{\alpha\dot{\beta}}\,P_{\mu}~,
\\[10pt]
[L_{\mu\nu},\,Q_{\alpha}]\!\!& =\!\!&-({\sigma}_{\mu\nu})_{\alpha}^{\;\beta}\,Q_{\beta},
\qquad\quad\; [L_{\mu\nu},\, \bar{Q}_{\dot{\alpha}}]\!\!&= \!\!&\bar{Q}_{\dot{\beta}}\,
({\bar{\sigma}}_{\mu\nu})_{\;\dot{\alpha}}^{\dot{\beta}}~,
\\[10pt]
[P_\mu^{},\,Q_{\alpha}]\!\!& =\!\!&0~,\qquad\qquad\qquad\qquad\;
[P_\mu^{},\,\bar{Q}_{\dot{\alpha}}]\!\!&=\!\!&0~,
\end{array}%
\end{eqnarray}
where by using the ordinary Pauli matrices $\sigma^{i}$ ($i=1,2,3$)  one sets:
$\sigma^{\mu}=(\mathbb{I}_{2},\, \sigma^i)$ and $\bar\sigma^{\mu}=(\mathbb{I}_{2},
-\sigma^i)$ for the Poincar\'{e} case and $\sigma^{\mu}=(i\mathbb{I}_{2},\, \sigma^i)$
and $\bar\sigma^{\mu}=(i\mathbb{I}_{2},\,-\sigma^i)$ for the Euclidean case. Following
typical spinorial notation one reads $(\sigma^{\mu})_{\alpha\dot\beta}$ and
$(\bar\sigma^{\mu})^{\dot\alpha\beta}$; the matrices $\sigma_{\mu\nu}=\frac{i}{4}
(\sigma_\mu\bar{\sigma}_\nu-\sigma_\nu\bar{\sigma}_\mu)$ and $\bar{\sigma}_{\mu\nu}=
\frac{i}{4}(\bar{\sigma}_\mu \sigma_\nu-\bar{\sigma}_\nu\sigma_\mu)$ are the generators
of $D=4$ Lorentz and Euclidean algebras given in two fundamental two-dimensional
spinorial representations. Moreover the antiinvolutions $(^*)$ in (\ref{se2}) are lifted
from Poincar\'{e} and Euclidean Lie algebras to their superextensions as follows:
\begin{eqnarray}\label{se4}
Q_{\alpha}^{*}\!\!&=\!\!&\bar{Q}_{\dot{\alpha}},\qquad\quad\;
\bar{Q}_{\dot{\alpha}}^{*}\;=\;Q_{\alpha}\qquad\;\;{\rm for}\;\;\mathcal{P}(3,1|1)~,
\\[10pt]\label{se5}
Q_{\alpha}^{*}\!\!&=\!\!&\varepsilon_{\alpha\beta}Q_{\beta},\qquad
\bar{Q}_{\dot{\alpha}}^{*}\;=\;\varepsilon_{\dot{\alpha}\dot{\beta}}
\bar{Q}_{\dot{\beta}}\quad\;{\rm for}\;\;\mathcal{E}(4|1)~,
\end{eqnarray}
where $\varepsilon_{\alpha\alpha}=\varepsilon_{\dot{\alpha}\dot{\alpha}}=0$,
$\varepsilon_{12}=-\varepsilon_{21}=-\varepsilon_{\dot{1}\dot{2}}=
\varepsilon_{\dot{2}\dot{1}}=1$. It should be noted that antiinvolution $(^*)$ in
(\ref{se4}) is the antilinear antiautomorphism of second order (conjugation) $(x^*)^*=x$
for $\forall x\in\mathcal{P}(3,1|1)$ and they reduce by half the number of independent
Poincar\'{e} supercharges. The constraints (\ref{se4}) together with (\ref{se2}) define
the reality condition for $N=1$ Poincar\'{e} superalgebra. The star operation $(^*)$ in
the fermionic sector of the supercharges (\ref{se3}) for the Euclidean superalgebra, due
to the relation $(\varepsilon_{\alpha\beta})^2=(\varepsilon_{\dot{\alpha}\dot{\beta}})^2=
-\mathbb{I}_2$, is the antilinear antiautomorphism of fourth order called
pseudoconjugation (for fermionic generators $(Q^*)^*=-Q$) and two pairs of Euclidean
supercharges $Q_{\alpha}$ and $\bar{Q}_{\dot{\alpha}}$ should be treated as independent.
In particular, the pseudoconjugation (\ref{se5}) can not be implemented by Hermitian
conjugation in Hilbert space.

We will also use below $\mathfrak{so}(3)$ physical basis in the Lie algebras
$\mathfrak{o}(3,1)$ and $\mathfrak{o}(4)$. Namely, we put
\begin{eqnarray}\label{se6}
M_i\!\!&:=\!\!&\epsilon_{ijk}L_{jk}~,\qquad N_i\;:=\;L_{0i}\quad (i,j=1,2,3).
\end{eqnarray}
In the terms of these elements the defining relations (\ref{se1}), (\ref{se3}) for
$\mathcal{P}(3,1|1)$ and $\mathcal{E}(4|1)$ take the form for the bosonic sector
\begin{eqnarray}\label{se7}
\begin{array}{rcccccl}
[M_i,\,M_j]\!\!&=\!\!&i\varepsilon_{ijk}M_k~,\qquad[M_i,\,N_j]\!\!&=\!\!&i
\varepsilon_{ijk}N_k~, \qquad[N_i,\,N_j]\!\!&=\!\!&\xi i\varepsilon_{ijk}M_k,
\\[10pt]
\qquad[M_i,\,P_j]\!\!&=\!\!&i\varepsilon_{ijk}P_k~,\qquad\;[M_i,\,P_0]\!\!&=\!\!&0~,
\qquad\qquad\;\;[N_i,\,P_j]\!\!&=\!\!&-i\delta_{ij}\,P_0~,
\\[10pt]
[N_i,\,P_0]\!\!&=\!\!&\xi iP_i~,\qquad\qquad[P_\mu,\,P_\nu]\!\!&=\!\!&0~,
\qquad\qquad\qquad\qquad&&
%% \quad(\mu,\nu=0,1,2,3)~,
\end{array}%
\end{eqnarray}
where now parameter $\xi$ distinguishes the Euclidean (Poincar\'{e}) cases $\xi=1$
($\xi=-1$) and for fermionic sector
\begin{eqnarray}\label{se8}
\begin{array}{rcccl}
\{Q_{\alpha},\,Q_{\beta}\}\!\!&=\!\!& \{\bar{Q}_{\dot{\alpha}},\,
\bar{Q}_{\dot{\beta}}\}\;=\;0,\qquad\{Q_{\alpha},\,\bar{Q}_{\dot{\beta}}\}\!\!&=\!\!&
2(\sigma^{\mu})_{\alpha\dot{\beta}}\,P_{\mu}~,
\\[12pt]
[M_{i},\,Q_{\alpha}]\!\!& =\!\!&\displaystyle -\frac{1}{2}
(\sigma^{i})_{\alpha}^{\;\beta}\,Q_{\beta},\qquad\qquad
[N_{i},\,Q_{\alpha}]\!\!&=\!\!&\displaystyle\chi\frac{1}{2}\,
(\sigma^{i})_{\alpha}^{\;\beta}\;Q_{\beta}~,
\\[12pt]
[M_{i},\,\bar{Q}_{\dot{\alpha}}]\!\!& =\!\!&\displaystyle\frac{1}{2}
\bar{Q}_{\dot{\beta}}\,(\sigma^{i})_{\;\alpha}^{\beta}\,,\qquad\qquad
[N_{i},\,\bar{Q}_{\dot{\alpha}}]\!\!&=\!\!&\displaystyle\chi\frac{1}{2}
\bar{Q}_{\dot{\beta}}\,(\sigma^{i})_{\;\alpha}^{\beta}~,
\\[16pt]
[P_\mu^{},\,Q_{\alpha}]\!\!&
=\!\!&[P_\mu^{},\,\bar{Q}_{\dot{\alpha}}]\;=\;0~,\qquad\qquad\qquad
\end{array}%
\end{eqnarray}
where  $\chi=1$ for the super-Euclidean case and  $\chi=-i$ for the super-Poincar\'{e}
case. It is well-known that we can pass in the relations (\ref{se7}), (\ref{se8}) from
the super-Poincar\'{e} to the super-Euclidean case by the replacement
\begin{eqnarray}\label{se9}
P_{i}\rightarrow P_{i}~,\qquad P_{0}\rightarrow iP_{0}~,\qquad M_{i}\rightarrow
M_{i}~,\qquad N_{i}\rightarrow iN_{i}~.
\end{eqnarray} %%
The replacement (\ref{se9}) can be justified by the change $x_{0}\rightarrow ix_{0}$ of
physical real time (Poincare case) into the purely imaginary Euclidean time.

\setcounter{equation}{0}
\section{Classical $r$-matrices of $N=1$ Poincar\'{e} and Euclidean superalgebras}
In this paper we will use for construction of covariant deformations of the
super-Minkowski and super-Euclidean  space-time  the quantum deformations obtained by
twist factors of the corresponding superalgebras. Such quantum deformations are
classified  by the classical supersymmetric $r$-matrices. Since the considered
superalgebras contain the Poincar\'{e} and Euclidean Lie algebras as subalgebras we shall
firstly consider the classical $r$-matrices for these Lie algebras.

\noindent{\large\textit{(1) Non-supersymmetric case}}. For the Poincar\'{e} algebra the
classical $r$-matrices were almost completely classified already some time ago by S.
Zakrzewski in \cite{Z94} for the Lorentz algebra and in \cite{Z97} for the Poincar\'{e}
algebra. We shall briefly remind these results.

It  was shown in \cite{Z97} that each classical $r$-matrix, $r\in\mathcal{P}(3,1)\wedge
\mathcal{P}(3,1)$, has a decomposition
\begin{eqnarray}\label{cr1}
r\!\!&=\!\!&a+b+c~,
\end{eqnarray}
where $a\in\mathbf{P}\wedge\mathbf{P}$, $b\in\mathbf{P}\wedge\mathfrak{o}(3,1)$,
$c\in\mathfrak{o}(3,1)\wedge\mathfrak{o}(3,1)$. The terms $a$, $b$, $c$ of $r$  satisfy
the following relations:
\begin{eqnarray}\label{cr2}
\begin{array}{rcl} %%
[[c,c]]\!\!&=\!\!&0~,
\\[10pt]
[[b,c]]\!\!&=\!\!&0~,
\\[10pt]
2[[a,c]]+[[b,b]]\!\!&=\!\!&t\Omega\quad (t\in \mathbb{R},\;\Omega\neq0)~,
\\[10pt]
[[a,b]]\!\!&=\!\!&0~,
\end{array}
\end{eqnarray}
where $[[\cdot,\cdot]]$ means the Schouten bracket, and $\Omega$ is $\mathfrak{g}$-%
invariant element, $\Omega\in(\stackrel{3}\wedge \mathfrak{g})_{\mathfrak{g}}$
($\mathfrak{g}=\mathcal{P}(3,1))$. A complete list of the classical $r$-matrices was
found for the case $c\neq0$ and as well for the case $c=0$, $t=0$ ; classification of the
$r$-matrices for the case $c=0$, $t\neq0$ is still not complete.
The results of \cite{Z97} are presented in {\bf Table 1}:\\[2pt]

{\small
\begin{tabular}{ccccc}
\hline $c$ & $b$ & $a$ & $\#$ & ${\cal N}$\\ %%
\hline $\gamma h'\wedge h$ & $0$ & $\alpha P_{+}\wedge P_{-}+\tilde{\alpha}P_{1}\wedge
P_{2}$ & $2$ & $1$\\
\hline $\gamma e'_{+}\wedge e_{+}$ & $\beta_{1}b_{P_{+}}^{}+\beta_{2}P_{+}\wedge h'$ &
$0$ & $1$ & $2$\\
$$ & $\beta_{1} b_{P_{+}}^{}$ & $\alpha P_{+}\wedge P_{1}$ & $1$ & $3$\\
$$ & $\gamma\beta_{1}(P_{1}\wedge e_{+}+P_{2}\wedge e'_{+})$ & $P_{+}\wedge(\alpha_{1}
P_{1}\!+\alpha_{2}P_{2})-\gamma\beta_{1}^2P_{1}\wedge P_{2}$ & $2$ & $4$\\
%% $$ & $$ & $-\gamma\beta_{1}^2P_{1}\wedge P_{2}$ & $$ & $$\\
\hline $\gamma(h\wedge e_{+}$ & $$ & $$ & $$ & $$\\
$-h'\wedge e'_{+})$ & $0$ & $0$ & $1$ & $5$\\
$+\gamma_{1}e'_{+}\wedge e_{+}$ & $$ & $$ & $$ & $$\\
\hline $\gamma h\wedge e_{+}$ & $\beta_{1}b_{P_{2}}^{}+\beta_{2}P_{2}\wedge e_{+}$ & $0$
& $1$ & $6$\\
\hline $0$ & $\beta_{1}b_{P_{+}}^{}+\beta_{2}P_{+}\wedge h'$ & $0$ & $1$ & $7$\\
$$ & $\beta_{1}b_{P_{+}}^{}+\beta_{2}P_{+}\wedge e_{+}$ & $0$ & $1$ & $8$\\
$$ & $P_{1}\wedge(\beta_{1}e_{+}+\beta_{2}e'_{+})\,+$ & $\alpha P_{+}\wedge P_{2}$ &
$2$ & $9$\\
$$ & $\beta_{1}P_{+}\wedge(h+\sigma e_{+}),\;\sigma=0,\pm1$ & $$ & $$ & $$\\
$$ & $\beta_{1}(P_{1}\wedge e'_{+}+P_{+}\wedge e_{+})$ & $\alpha_{1}P_{-}\wedge
P_{1}+\alpha_{2} P_{+}\wedge P_{2}$ & $2$ & $10$\\
$$ & $\beta_{1} P_{2}\wedge e_{+}$ & $\alpha_{1}P_{+}\wedge P_{1}+\alpha_{2}
P_{-}\wedge P_{2}$ & $1$ & $11$\\
$$ & $\beta_{1} P_{+}\wedge e_{+}$ & $P_{-}\!\wedge(\alpha P_{+}\!+\!\alpha_{1}P_{1}\!+\!
\alpha_{2}P_{2})\!+
\tilde{\alpha} P_{+}\!\wedge P_{2}$ & $3$ & $12$\\
$$ & $\beta_{1} P_{0}\wedge h'$&$\alpha_{1}P_{0}\wedge P_{3}+\alpha_{2}P_{1}\wedge P_{2}$&
$2$ & $13$\\
$$ & $\beta_{1} P_{3}\wedge h'$&$\alpha_{1}P_{0}\wedge P_{3}+\alpha_{2}P_{1}\wedge P_{2}$&
$2$ & $14$\\
$$ & $\beta_{1} P_{+}\wedge h'$&$\alpha_{1}P_{0}\wedge P_{3}+\alpha_{2}P_{1}\wedge P_{2}$&
$1$ & $15$\\
$$ & $\beta_{1} P_{1}\wedge h$&$\alpha_{1}P_{0}\wedge P_{3}+\alpha_{2}P_{1}\wedge P_{2}$ &
$2$ & $16$\\
$$ & $\beta_{1} P_{+}\wedge h$&$\alpha P_{1}\wedge P_{2}+\alpha_{1}P_{+}\wedge P_{1}$ &
$1$ & $17$\\
$$ & $P_{+}\wedge(\beta_{1} h+\beta_{2} h')$&$\alpha_{1} P_{1}\wedge P_{2}$ & $1$ & $18$\\
\cline{2-5}
$$ & $0$ & $\alpha_{1} P_{1}\wedge P_{+}$ & $0$ & $19$\\
$$ & $$ & $\alpha_{1} P_{1}\wedge P_{2}$ & $0$ & $20$\\
$$ & $$ & $\alpha_{1} P_{0}\wedge P_{3}+\alpha_{2}P_{1}\wedge P_{2}$ & $1$ & $21$\\
\hline
\end{tabular}}\\[10pt]
where $P_{\pm}=P_{0}\pm P_{3}$ and $b_{P_{+}}^{}$, $b_{P_{2}}^{}$ are given by the
expressions:
\begin{eqnarray}\label{cr3}
\begin{array}{rcl}
b_{P_{+}}^{}\!\!&=\!\!&P_{1}\wedge e_{+}-P_{2}\wedge e'_{+}+P_{+}\wedge h~,
\\[6pt]
b_{P_{2}}^{}\!\!&=\!\!&2P_{1}\wedge h'+P_{-}\wedge e'_{+}-P_{+}\wedge e'_{-}~.
\end{array} %
\end{eqnarray}
\noindent The generators $e_{\pm},h,e'_{\pm},h'$ in Table 1 describe the canonical
(mathematic) basis of the Lorentz Lie algebra $\mathfrak{o}(3,1)$ which is obtained by
realification of $\mathfrak{sl}(2, \mathbb{C})$ (see \cite{Z97, BLT08}) and satisfy
the following non-vanishing commutation relations: %
\begin{eqnarray}\label{cr4}
\begin{array}{rcccl}
[h,\,e_{\pm}^{}]\!\!&=\!\!&\pm e_{\pm}^{}~,\qquad [e_{+}^{},\,e_{-}^{}]\!\!&=\!\!&2h~,
\\[6pt]
[h,\,e'_{\pm}]\!\!&=\!\!&\pm e'_{\pm}~,\qquad [h',\,e_{\pm}]\!\!&=\!\!&\pm
e'_{\pm}~,\quad\;\; [e_{\pm}^{},\,e'_{\mp}]\;=\;\pm2h'~,
\\[6pt]
[h',\,e'_{\pm}]\!\!&=\!\!&\mp e_{\pm}^{}~,\qquad [e'_{+},\,e'_{-}]\!\!&=\!\!&-2h~.
\end{array}
\end{eqnarray} %%
Table 1 lists 21 cases labeled by the number ${\cal N}$ in the last column. The forth
column (labeled by $\#$) indicates a maximal number of {\it independent} parameters
defining deformations. This number is in all cases smaller than the number of parameters
actually used in the Table 1. Following \cite{T07}, we introduced an additional parameter
$\gamma$ in the component $c$ (in the cases ${\cal N}=2,\ldots,6$), a parameter
$\beta_{1}^{}$ in the component $b$ (in the cases ${\cal N}=7,\ldots,18$) and a parameter
$\alpha$ in the component $a$ (in the cases ${\cal N}=19,20,21$)\footnote{In the original
paper by S. Zakrzewski \cite{Z97} all these additional parameters are equal to 1,
therefore we should assume in Table 1 that they are not equal to zero.}. The maximal
numbers of independent parameters can be calculated using of automorphisms of the
Poincar\'{e} algebra ${\mathcal{P}}(3,1)$ (for details see \cite{Z97}).

Important point in our consideration is the property that the relations (\ref{cr4}) can
describe Lorentz $\mathfrak{o}(3,1)$ as well as the Euclidean $\mathfrak{o}(4)$ algebra.

The canonical generators $h,h',e_{\pm},e'_{\pm}$ are related with the physical generators
(\ref{se6}), (\ref{se7}) of the Lorentz and Euclidean algebra by the relations:
\begin{eqnarray}\label{cr5}
\begin{array}{rcccl}
h\!\!&=\!\!&\chi N_3~,\qquad e_{\pm}\!\!&=\!\!&(\chi N_1\pm iM_2)~,
\\[6pt]
h'\!\!&=\!\!&iM_3~,\qquad e'_{\pm}\!\!&=\!\!&(iM_1\mp\chi N_2)~.
\end{array} %%
\end{eqnarray}
where as before $\chi=-i$ for Lorentz algebra $\mathfrak{o}(3,1)$, and $\chi=1$ for
Euclidean $\mathfrak{o}(4)$. It follows from (\ref{cr5}) that in Poincar\'{e} case the
canonical basis is anti-Hermitian, i.e. %%
\begin{eqnarray}\label{cr6} %%
x^*\!\!&=\!\!&-x\qquad (\forall x\in \{e_{\pm},h,e'_{\pm},h'\})~,
\end{eqnarray} %%
but the generators (\ref{cr5}) for the Euclidean case have different reality properties
with respect to the conjugation (\ref{se2}) in $\mathfrak{o}(4)$, namely
\begin{eqnarray}\label{cr7}
\begin{array}{rcccl}
e_{\pm}^*\!\!&=\!\!&e_{\mp}~,\qquad h^*\!\!&=\!\!&h~,\qquad{e'_{\pm}}^*\;=\;-e'_{\mp}
\qquad {h'}^*\;=\;-h'~,~.
\end{array} %%
\end{eqnarray}
If we introduce in the Poincar\'{e} case all the four-momenta generators $P_{1}$,
$P_{2}$, $P_{\pm}$ anti-Hermitian (purely imaginary), we see that all the classical
$r$-matrices with real deformation parameters in Table 1 for the Poincar\'{e} Lie algebra
are Hermitian ($r^*=r$). \\
It should be noted that for all (but ${\cal N}=6,12$) classical $r$-matrices\footnote{It
should be noted that in the paper \cite{Z97} there is a misprint for the case ${\cal
N}=12$ which is corrected here.} of Table 1 corresponding quantum deformations of
$\mathcal{P}(3,1)$, described by twists, were constructed in \cite{T07}.

It appears however that Table 1 can be used  as well for Euclidean Lie algebra because
all $r$-matrices of Table 1 are the classical $r$-matrices for complexified
$\mathcal{E}(4)$. Indeed, if we replace in the formulas (\ref{cr5}) generators $M_i$ and
$N_i$ of $\mathfrak{o}(3,1)$ by the generators of $\mathfrak{o}(4)$ and put $\tau=1$ then
the new generators will satisfy the same relations (\ref{cr4}). Moreover the commutation
relations of the rotation generators with the four momenta $P_{\mu}$ also remain
unchanged if we replace $P_{\pm}=P_{0}\pm P_{3}$ for Lorentzian metric by
$P_{\pm}=iP_{0}\pm P_{3}$ for (compare for (\ref{se9})) Euclidean one.  Therefore all the
$r$-matrices of Table 1 will satisfy the classical Yang-Baxter equation in Euclidean case
however due to complex values of $P_\pm$ and the reality constraints (\ref{cr7}) (complex
$e_\pm$ and $e^\prime_\pm$) we observe that only six classical $r$-matrices from Table 1
with ${\cal N}=1,13,14,16,20,21$ describe in Euclidean case the Hermitian classical
$r$-matrices. Additional two Euclidean real classical $r$-matrices can be described by
the formulae from Table 1 with ${\cal N}=15$ and ${\cal N}=19$ provided we keep the
formula $P_\pm=P_0\pm P_3$ as well in Euclidean case.

\noindent{\large\textit{(2) Supersymmetric case}}. Firstly we observe that all classical
$o(3,1)$ ($o(4)$) $r$-matrices satisfying homogeneous YBE are as well the $r$-matrices
for the Poincar\'{e} (Euclidean) super-algebra and the corresponding twists of the
bosonic subalgebra can be used for the deformation of full superalgebra. Further we shall
be interested in  the supersymmetric extension of classical $r$-matrices of the
super-Poincar\'{e} algebra ${\mathcal P}(3,1|1)$, which contain the supercharges
$Q_{\alpha}$ ($\alpha=1,2$) and $\bar{Q}_{\dot{\alpha}}$ ($\dot{\alpha}=\dot{1},\dot{2}$)
and will consider also their Euclidean counterparts. It should be noted that there is not
known any classification of such $r$-matrices in spirit of the classification done by
S.~Zakrzewski.  It turns out however that by explicit calculations it is possible to
extend the Zakrzewski's classification to the Poincar\'{e} and Euclidean superalgebras by
an addition of  terms depending on supercharges. The superextensions of Zakrzewski's
$r$-matrices are presented in {\bf Table 2}:
\\[10pt]

%%{\tiny
{\scriptsize
%%{\small
\begin{tabular}{cccccc}
\hline\smallskip $c$ & $b$ & $a$ & $s$ & ${\cal N}$\par\\
\hline$$$$$$$$$$\\
\par$\gamma h'\wedge h$ & $0$ & $\alpha P_{+}\wedge P_{-}\!+\tilde{\alpha}P_{1}
\wedge P_{2}$ & $\eta Q_{2}\wedge Q_{1}$ & $1$\smallskip\\

\hline$$$$$$$$$$\\ $\gamma e'_{+}\wedge e_{+}$ & $\beta_{1}b_{P_{+}}^{}+\beta_{2}P_{+}\wedge h'$ &
$0$ & $\beta_{1}\bar{Q}_{\dot 1}\wedge Q_{1}$ & $2$\smallskip\\
$ $ & $\beta_{1} b_{P_{+}}^{}$ & $\alpha P_{+}\wedge P_{1}$ &
$\beta_{1}\bar{Q}_{\dot 1}\!\wedge \!Q_{1}\!+\!\eta Q_{1}\!\wedge\!Q_{1}$ & $3$\\
$$ & $\gamma\beta_{1}(P_{1}\wedge e_{+}+P_{2}\wedge e'_{+})$ & $P_{+}\wedge(\alpha_{1}P_{1}\!+
\alpha_{2}P_{2})%% -\gamma\beta_{1}^2P_{1}\wedge P_{2}
$ & $\eta Q_{1}\wedge Q_{1}$ & $4$\\
$$ & $$ & $-\gamma\beta_{1}^2P_{1}\wedge P_{2}$ & $$ & $$ & $$\smallskip\\
\hline%\hline\medskip
$\gamma(h\wedge e_{+}$ & $$ & $$ & $$ & $$ & $$\\
$-h'\wedge e'_{+})$ & $0$ & $0$ & $0$ & $5$\\
$+\gamma_{1}e'_{+}\wedge e_{+}$ & $$ & $$ & $$ & $$\par\\

\hline$$$$$$$$$$\\
\par$\gamma h\wedge e_{+}$ & $\beta_{1}b_{P_{2}}^{}+\beta_{2}P_{2}\wedge e_{+}$ & $0$
& $i\beta_{1}(Q_{1}+\bar{Q}_{\dot 1})\wedge(Q_{2}-\bar{Q}_{\dot 2})$ & $6$\par\\

\hline$$$$$$$$$$\\
$0$ & $\beta_{1}b_{P_{+}}^{}+\beta_{2}P_{+}\wedge h'$ & $0$ &
$\beta_{1}\bar{Q}_{\dot 1}\wedge Q_{1}$ & $7$\\
$$ & $\beta_{1}b_{P_{+}}^{}+\beta_{2}P_{+}\wedge e_{+}$ & $0$ &
$\beta_{1}\bar{Q}_{\dot 1}\!\wedge\!Q_{1}\!+\!\eta Q_{1}\!\wedge\!Q_{1}$ & $8$\\
$$ & $P_{1}\wedge(\beta_{1}e_{+}+\beta_{2}e'_{+})+$ & $\alpha P_{+}\wedge P_{2}$ &
$\beta_{1}\bar{Q}_{\dot 1}\!\wedge\!Q_{1}\!+\!\eta Q_{1}\!\wedge\! Q_{1}$ & $9$\\
$$ & $\beta_{1}P_{+}\!\wedge\!(h\!+\!\chi e_{+}),\chi=0,\pm1$ & $$ & $$ & $$ & $$\\
$$ & $\beta_{1}(P_{1}\wedge e'_{+}+P_{+}\wedge e_{+})$ & $\alpha_{1}P_{-}\wedge
P_{1}+\alpha_{2} P_{+}\wedge P_{2}$ & $\eta Q_{1}\wedge Q_{1}$ & $10$\\
$$ & $\beta_{1} P_{2}\wedge e_{+}$ & $\alpha_{1}P_{+}\wedge P_{1}+\alpha_{2}
P_{-}\wedge P_{2}$ & $\eta Q_{1}\wedge Q_{1}$ & $11$\\
$$ & $\beta_{1} P_{+}\wedge e_{+}$ & $P_{-}\!\wedge(\alpha P_{+}\!+\!\alpha_{1}P_{1})+
%%\alpha_{2}P_{2})+
%% \tilde{\alpha} P_{+}\!\wedge P_{2}
$ & $\eta Q_{1}^{}\wedge Q_{1}^{}$ & $12$\\
$$ & $$ & $\tilde{\alpha} P_{+}\!\wedge P_{2}$ & $$ & $$ & $$\\
$$ & $\beta_{1}P_{0}\wedge h'$ & $\alpha_{1}P_{0}\wedge P_{3}+\alpha_{2}P_{1}\wedge P_{2}$&
$\eta Q_{2}\wedge Q_{1}$ &$13$\\
$$ & $\beta_{1}P_{3}\wedge h'$ & $\alpha_{1}P_{0}\wedge P_{3}+\alpha_{2}P_{1}\wedge P_{2}$&
$\eta Q_{2}\wedge Q_{1}$ & $14$\\
$$ & $\beta_{1}P_{+}\wedge h'$ & $\alpha_{1}P_{0}\wedge P_{3}+\alpha_{2}P_{1}\wedge P_{2}$ &
$\eta Q_{2}\wedge Q_{1}$ & $15$\\
$$ & $\beta_{1} P_{1}\wedge h$ & $\alpha_{1}P_{0}\wedge P_{3}+\alpha_{2}P_{1}\wedge P_{2}$ &
$\eta Q_{2}\!\wedge\! Q_{1}$ & $16$\\
$$ & $\beta_{1} P_{+}\wedge h$ & $\alpha_2 P_{1}\wedge P_{2}+\alpha_{1}P_{+}\wedge P_{1}$ &
$\beta_{1}\bar{Q}_{\dot 1}\!\wedge\! Q_{1}\!+\!\eta Q_{1}\!\wedge\!Q_{1}$ & $17$\\
$$ & $P_{+}\wedge(\beta_{1}h+\beta_{2} h')$ & $\alpha_{2} P_{1}\wedge P_{2}$ &
$\eta Q_{2}\wedge Q_{1}$ & $18$\par\\
\cline{2-6}$$$$$$$$\\
%\hline$$$$$$$$$$\\
$$ & $0$ & $\alpha_{1}P_{1}\wedge P_{+}$ & $\eta^{\alpha\beta}_{}Q_{\alpha}\wedge
Q_{\beta}$ & $19$\\
$$ & $$ & $\alpha_{2}P_{1}\wedge P_{2}$ & $\eta^{\alpha\beta}_{}Q_{\alpha}\wedge
Q_{\beta}$ & $20$\\
$$ & $$ & $\alpha_{1} P_{0}\wedge P_{3}+\alpha_{2}P_{1}\wedge P_{2}$ &
$\eta^{\alpha\beta}_{}Q_{\alpha}\wedge Q_{\beta}$ & $21$\\

\hline
\end{tabular}}\\[10pt]

\noindent These supersymmetric $r$-matrices can be presented as a sum of subordinated
$r$-matrices which are of super-Abelian and super-Jordanian types. The subordination
enables us to construct a correct sequence of quantizations and to obtain the
corresponding twists describing the quantum deformations. These twists are in general
case complex super-extensions of the twists obtained in \cite{T07}.

Let us now select out of the supersymmetric $r$-matrices $r_{susy}$ from Table 2
$$
r_{susy}\,=\,r+s\,=\,a+b+c+s
$$
the ones that are self-conjugate respectively under the Poincar\'{e} conjugation (see
(\ref{se2}) and (\ref{se4})) and Euclidean pseudo-conjugation (see (\ref{se2}) and
(\ref{se5})).\\
\noindent{\large\textit{(i) Real classical super-Poincar\'{e} $r$-matrices}}. Following
\cite{Z97}, all classical Poincar\'{e} $r$-matrices from Table 1 are real (Hermitian). It
appears however that only seven out of 21 cases are real after  supersymmetrization,
namely
\begin{eqnarray}\label{cr8}
\begin{array}{rcccccl}
1).\;\;{\cal N}\!\!&=\!\!&2~,\qquad 2).\;\; {\cal N}\!\!&=\!\!&3\;\;{\rm
for}\;\;\eta=0~,\qquad\;\; 3).\;\;{\cal N}\!\!&=\!\!&6~,
\\[7pt]
4).\;\;{\cal N}\!\!&=\!\!&7~,\qquad 5).\;\;{\cal N}\!\!&= \!\!&8,9\;\;{\rm
for}\;\;\eta=0~,\qquad 6).\;\;{\cal N}\!\!&=\!\!&17\;\;{\rm for}\;\;\eta=0~.
\end{array}%
\end{eqnarray}
\noindent{\large\textit{(ii) Real (self-conjugate) classical super-Euclidean
$r$-matrices}}. Let us observe firstly that only 8 out of 21 cases (${\cal N}=1, 13-16,
19-21$) listed in the Table 1 describe real classical Euclidean $r$-matrices. Out of them
the following ones provide  self-pseudoconjugate super-Euclidean $r$-matrices:
\begin{eqnarray}\label{i6}
\begin{array}{rcl}
1).\;\;{\cal N}\!\!&=\!\!&1~,\quad 2).\;\; {\cal N}\;=\;13-16\;\;({\rm
with}\;\;P_{+}=P_{0}+P_{3}\;\;{\rm for}\;\;{\cal N}=15)~,
\\[7pt]
3).\;\;{\cal N}\!\!&=\!\!&19-21\;\;({\rm with}\;\;P_{+}=P_{0}+P_{3}\;\;{\rm for}\;\;{\cal
N}=19)~.
\end{array}%
\end{eqnarray}
The cases ${\cal N}=13-16$ will be considered separately  more in details below. The
cases ${\cal N}=19-21$ describe three basic superextension of the canonical (Moyal--Weyl)
deformations which were considered by several authors \cite{KS}--\cite{IS}.

Finally we would like to mention that our classification technique of Euclidean
$r$-matrices and corresponding supersymmetric $r$-matrices has one limitation: it is
obtained by the ''Euclideization'' of the Zakrzewski Table 1 for Poincar\'{e}
$r$-matrices. It is however quite possible that there are Euclidean $r$-matrices with
self-pseudoconjugate supersymmetric extension which are not corresponding to the
$r$-matrices presented in Table 1.

\setcounter{equation}{0}
\section{Twist deformations of (super)space-time}
{\large\textit{(1) Twisted deformations of Poincar\'{e} and Euclidean algebras and
deformed Minkowski and Euclidean space-times}}. Let us consider firstly
non-supersymmetric cases. At the beginning of this section the Poincar\'{e} and Euclidean
cases will be simultaneously analyzed and therefore we introduce the following unified
denotations: deformed Poincar\'{e} algebra and Minkowski space-time are denoted by
$U_{\kappa}(\mathcal{A}_1):=U_{\kappa}(\mathcal{P}(3,1))$ and $U_{\kappa}(V_{1})
:=U_{\kappa}(\mathcal{M}(3,1))$; analogously deformed Euclidean algebra and Euclidean
space-time are denoted by $U_{\kappa}(\mathcal{A}_2):=U_{\kappa}(\mathcal{E}(4))$ and
$U_{\kappa}(V_2):=U_{\kappa}(E(4))$, where $\kappa$ describes the mass-like deformation
parameter.

Most of the deformed space-times considered in the literature can be described by
constant and linear values of the commutator of quantum space-time variables $%
\hat x^\mu\in V_i$ ($\mu=0,1,2,3;\, i=1,2$):
\begin{eqnarray}\label{i1}
[\hat{x}^{\mu},\,\hat{x}^{\nu}]\!\!&=\!\!&\frac{i}{\kappa^2}\Theta ^{\mu\nu}_{}
(\kappa\hat{x})\,=\,\frac{i}{\kappa^2}\left({\vartheta}^{\mu\nu}_{}+
\kappa{\vartheta}^{\mu\nu}_{\lambda}\hat{x}^\lambda+\dots\right)
\end{eqnarray}
where the tensor $\Theta^{\mu\nu}_{}(\kappa\hat{x})\}$ is determined by the dimensionless
set of constant parameter $\{\vartheta^{\mu\nu}\}$, $\{\vartheta^{\mu\nu}_{\lambda}\}$,
etc. A large class of deformations (\ref{i1}) can be interpreted as covariant under
twisted $\mathcal{A}_i$ symmetries. If $\Theta^{\mu\nu}(\kappa\hat{x})=
\Theta^{\mu\nu}(0)\equiv \vartheta^{\mu\nu} $ (a case of the simplest canonical or
Moyal--Weyl deformation) the corresponding Abelian twist is the following \cite{Oeckl} -
\cite{Wess}
\begin{eqnarray}\label{i2}
F_{0}^{}\!\!&=\!\!&\exp\Bigl({\frac{i}{2\kappa^2}}\vartheta^{\mu\nu}_{}\,P_\mu\wedge
P_\nu\Bigr).
\end{eqnarray}
The quantum-deformed Hopf algebra $\hat{U}_\kappa(\mathcal{A}_i)$ acts on the enveloping
algebra $U_{\kappa}(V_i)$ of the corresponding quantum-deformed spacetime (see
(\ref{i1})) as its representation (Hopf-algebra module). If $\hat{g}\in U_\kappa(A_i)$
and $\hat{x},\hat{y}\in U_\kappa(V_i))$ the Hopf algebraic action has the property
\begin{eqnarray}\label{i3}
\hat{g}\vartriangleright(\hat{x}\hat{y})\!\!&=\!\!&(\hat{g}%
_{(1)}\vartriangleright \hat{x})(\hat{g}_{(2)}\vartriangleright\hat{y})~,
\end{eqnarray}
where we use the standard Sweedler's notation $\Delta(\hat{g})=\hat{g}_{(1)}\otimes\hat
g_{(2)}$. The deformation of classical Hopf algebra $\hat{U}(\mathcal{A}_i)$ by twist $F$
does not modify the algebraic sector described by (\ref{se1}) but it changes the
primitive coproduct $\Delta_0(\hat{g})=\hat{g}\otimes1+1\otimes\hat{g}$
($\hat{g}\in\mathcal{A}_i$) and antipode $S_0(\hat g)=-\hat g$ as follows \cite{Dr}:
\begin{equation}\label{i4}
\Delta^{(F)}F(\hat{g})=F\Delta _0(\hat{g})F^{-1},\qquad S_F(\hat{g})=uS_0(\hat{g}
)u^{-1},
\end{equation}%
where $u=F_{(1)}S_0(F_{(2)})$ for $F=F_{(1)}\otimes F_{(1)}$.

If the algebra (\ref{i1}) is associated with twisted $\mathcal{A}_i$-symmetries defined
by a twist $F$ the multiplication of noncommutative coordinates $\hat{x}^{\mu}$ can be
isomorphically represented by suitable star multiplication of the commuting classical
coordinates
\begin{eqnarray} \label{i5}
f(\hat{x})\varphi(\hat{x})\overset{W}{\longrightarrow} f(x)\star\varphi(x)
\quad\Rightarrow\quad\hat{x}^{\mu}\hat{x}^{\nu}\overset{W}{\longrightarrow}
x^{\mu}\star x^{\nu}~,
\end{eqnarray}
where $W$ denotes the Weyl map, defined for the twist factor $F$ by the formula (see
\cite{Oeckl}--\cite{Kulish}) %%
\begin{eqnarray}\label{i6}
\begin{array}{rcl}
f(x)\star\varphi(x)\!\!&=\!\!&f(x)\star_{F^{}}\varphi(x)\,:=\,m\Bigl(F_{}^{-1}
\triangleright\bigl(f(x)\otimes\varphi(x)\bigr)\Bigr)\quad\Rightarrow
\\[7pt] %%
x^{\mu}\star x^{\nu}\!\!& =\!\!& x^{\mu}\star_{F^{}}x^{\nu}\;=\;m\bigl(F_{}^{-1}
\triangleright(x^{\mu}\otimes x^{\nu})\bigr)~,%
\end{array}%
\end{eqnarray}
If we insert $F=F_0$ (see (\ref{i2})) one obtains the relation (\ref{i1}) with
$\Theta^{\mu\nu}(\kappa\hat{x})=\vartheta^{\mu\nu}$
\begin{eqnarray}\label{i7}
[x^\mu, x^\nu]_{\star_{F_0}^{}}\!\!&\equiv\!\!& x^\mu\star_{F_0} x^\nu-x^\nu\star_{F_0}
x^\mu=\frac{\imath}{\kappa^2}\vartheta^{\mu\nu}
\end{eqnarray}
The last relation is quantum--covariant, i.e. one can show using (\ref{i3}) and
(\ref{i4}) that
\begin{eqnarray}\label{i8}
\hat{g}\triangleright\Bigl([x^\mu, x^\nu]_{\star_{F_0^{}}^{}}-\frac{i}{%
\kappa^2} \vartheta^{\mu\nu}_{}\Bigr)\;=\;0
\end{eqnarray}
where $\hat{g}\in U_{\kappa}(\mathcal{A}_i)$ and the action of symmetry generators is
obtained in the $\star$-product realization.

In this paper we shall consider other family of Abelian Poincar\'{e} (Euclidean) twists
which are described by the following general formula
\begin{eqnarray}  \label{i9}
\mathcal{F}\!\!&=\!\!&F_{1}F_{0}\;\;=\;\;\exp\Big(\frac{i}{2\kappa}
\vartheta^{\mu\nu\lambda}\,L_{\mu\nu}\wedge P_{\lambda}\Big)\exp\Bigl({\frac{i} {%
2\kappa^2}}\vartheta^{\mu\nu}_{}\,P_\mu\wedge P_\nu\Bigr)~,
\end{eqnarray}
It follows from the paper \cite{T07} (see Table 1 ${\cal N}=13-16$) there are four
inequivalent (i.e. modulo a Poincar\'{e} transformations) Abelian Poincar\'{e} twists of
this type which in terms of the generators (\ref{se6}) are defined as follows: %%
\begin{eqnarray}\label{i10}
\begin{array}{rcl}
\mathcal{F}_{a}\!\! & =\!\! & F_{1a}\,F_{0}\;\;=\;\;\exp{(i\beta
M_{3}\wedge P_{0})}\,F_{0}~,\quad \\[10pt]
\mathcal{F}_{b}\!\! & =\!\! & F_{1b}\,F_{0}\;\;=\;\;\exp{(i\beta
M_{3}\wedge P_{3})}\,F_{0}~, \\[10pt]
\mathcal{F}_{c}\!\! & =\!\! & F_{1c}\,F_{0}\;\;=\;\;\exp{(i\beta
M_{3}\wedge P_{+})}\,F_{0}~, \\[10pt]
\mathcal{F}_{d}\!\! & =\!\! & F_{1d}\,F_{0}\;\;=\;\;\exp{(i\beta
N_{3}\wedge P_{1})}\,F_{0}~,%
\end{array}%
\end{eqnarray}
where $P_{+}:=P_{0}+P_{3}$ and the fourmomentum-dependent factor $F_{0}$ (the same for
all four formulas) is given by \footnote{It corresponds to particular choice of the
parameter $\vartheta^{\mu\nu}$ in (\ref{i2}).}
\begin{eqnarray}\label{i11}
F_{0}\!\!&=\!\!&\exp{\bigl(i\alpha_{1}P_{3}\wedge P_{0}+i\alpha_{2}P_{2}\wedge
P_{1}\bigr)}~.
\end{eqnarray}

\noindent{\large\textit{(2) The examples of twisted deformations of Euclidean
superspaces}}. It follows from Section 3 that four classical $r$-matrices (see Table 2,
${\cal N}=13-16$) as well as the corresponding twists (\ref{i10}) are valid for both
cases of Poincar\'{e} and Euclidean symmetries (with the formula $P_{+}=P_{0}+P_{3}$
valid also in Euclidean case). In this paper we shall describe more in detail the
supersymmetric extension of four twists (\ref{i10}) (other examples of super-Poincar\'{e}
were considered in \cite{BLT08}). We shall provide the algebraic structure of the
corresponding deformed chiral superspaces.

The $r$-matrices generating of the twists (\ref{i10}) are given by the following
formulas: %%
\begin{eqnarray}\label{i12}
r_{w}^{}\!\!&=\!\!&r_{1w}^{}+r_{0}^{}\qquad (w=a,b,c,d)~,
\end{eqnarray}%%
where the indices $w=(a,b,c,d)$ label four twists $F_{\omega}$,
\begin{equation}\label{i13}
\begin{array}{rcccl}
r_{1a}^{}\!\!&=\!\!&\beta\,P_{0}\wedge M_{3} ~,\qquad\qquad r_{1b}^{}\!\!&=\!\!&
\beta\,P_{3}\wedge M_{3}~, %%
\\[10pt] %%
r_{1c}^{}\!\!&=\!\!&\beta \,P_{+}\wedge M_{3}~,\qquad\qquad r_{1d}^{}\!\!&=\!\!&
\beta\,P_{1}\wedge N_{3}~, %%
\\[10pt] %%
r_{0}^{}\!\!&=\!\!&\alpha _{1}^{}P_{0}\wedge P_{3}+\alpha _{2}^{}P_{1}\wedge P_{2}~. &&
\end{array}%%
\end{equation}%%
From Table 2 we get the following formulae for the supersymmetric extension $r_{ws}^{}$
of the classical $r$-matrices $r_{w}^{}$ and the corresponding supertwists
$\mathcal{F}_{ws}$\footnote{It should be stressed that the order of factors in the
formula (\ref{i15}) is important, because the Abelian $r$-matrices $r_{1w}^{}$ and
$r_{1w}^{}+r_{s}^{}$ are subordinate. One say that $r_{2}^{}$ is subordinated to
$r_{1}^{}$, $r_{1}^{}\succ r_{2}^{}$, if $[x\otimes 1+ 1\otimes x,\,r_{1}^{{}}]=0$
($\forall x\in\mathop{\rm Sup}(r_{2})$), where $\mathop{\rm Sup} (r_{2})$ is a support of
$r_{2}$ (see \cite{T07} for details).}:
\begin{eqnarray}\label{i14}
r_{ws}^{}\!\!&=\!\!&r_{w}^{}+r_{s}^{}\,=\,r_{1w}^{}+r_{0}^{}+r_{s}^{}\qquad(w=a,b,c,d)~,
\\[10pt] \label{i15} %%
\mathcal{F}_{ws}^{}\!\!&=\!\!&F_{1w}^{{}}\,F_{0}^{}\,F_{s}^{}\,=\,
\exp\Bigl(\frac{i}{\kappa}\,\tilde{r}_{1w}^{}\Bigr)\exp\Bigl(\frac{i}{\kappa^{2}}\,
\tilde{r}_{0}^{}\Bigr)\exp\Bigl(\frac{i}{\kappa}\,\tilde{r}_{s}^{}\Bigr)~,
\end{eqnarray}%%
where $\tilde{r}_{w}^{}=-r_{w}^{}$ and the term $\tilde{r}_{s}^{}=r_{s}^{}$ describing
supersymmetric extension is given as follows
\begin{eqnarray}\label{i16}
r_{s}^{}\!\!&=\!\!&\eta\,Q_{2}\wedge Q_{1}\,=\,\eta(Q_{2}\otimes Q_{1} +Q_{1}\otimes
Q_{2})
\end{eqnarray}%%
and $\eta$ is a deformation parameter.

The super-term $F_{s}$ in (\ref{i15}) for any value of $\eta$ is not invariant under the
involution (\ref{se4}) characterizing the Poincar\'{e} superalgebra, and leads to the
co-product, which does not satisfy the condition
\begin{eqnarray}\label{i17}
(\Delta^{(F)}(g))^*\!\!&=\!\!&\Delta^{(F)}(g^*)~.
\end{eqnarray}%%
Consequently the deformed chiral ($Q_{\alpha}$) and antichiral ($\bar{Q}_{\dot{\alpha}}$)
supercharges will not be complex conjugate to each other. Fortunately, if we consider the
Euclidean case and corresponding pseudoconjugation (\ref{se5}) then for real $\eta$ we
obtain the following invariance relation for $r$-matrices $r_{s}$ (as well as $r_{ws}$)
\begin{eqnarray}\label{i18}
r_{s}^{*}\!\!&=\!\!&\eta\,Q_{2}^*\wedge Q_{1}^*\,=\,r_{s}
\end{eqnarray}%%
expressing its the pseudoreality property. Furthermore the super-twist $\mathcal{F}_{ws}$
will be (pseudo)unitary
\begin{eqnarray}\label{i19}
(\mathcal{F}_{ws})^{*}\!\!&=\mathcal{F}_{ws}^{-1},
\end{eqnarray}%%
where the pseudoconjugation (\ref{se5}) is used. It can be shown that the new deformed
co-product $\Delta^{(\mathcal{F})}:=\mathcal{F}_{ws}^{}\Delta_{0} \mathcal{F}_{ws}^{-1}$
satisfies the relation (\ref{i17}). We see therefore that if we discard the possibility
of doubling of the number of supercharges in the deformation procedure the twist
deformations generated by the super-twists (\ref{i15}) should be only applied in the
Euclidean case (not for Poincar\'{e} superalgebra).

\setcounter{equation}{0}
\section{Twist-deformed Euclidean chiral superspace} %%
In order to obtain the $\star$-products (and also $\star$-commutators) of the superspace
coordinates with help of the formula (\ref{i6}) (where the twist $F$ is replaced by
supertwists (\ref{i15})) we need a realization of the $N=1$ super-Euclidean algebra,
$\mathfrak{o}(4|1)$, in terms of linear differential operators on superspace. In
accordance with \cite{FZW} there are in Minkowski as well as in Euclidean superspace
three differential superspace realizations: non-chiral, chiral and antichiral. The chiral
supersymmetric covariant derivatives, anticommuting with the supercharges ($\{Q_{\alpha},
D_{\beta}\}=\{Q_{\alpha},D_{\dot{\beta}}\}=\{\bar{Q}_{\dot{\alpha}},D_{\beta}\}=
\{\bar{Q}_{\dot{\alpha}},D_{\dot{\beta}}\}=0$) are defined as follows
\begin{eqnarray}\label{ecs1} %%
D_{\alpha}\!\!&=\!\!&\partial _{\alpha}-2(\sigma ^{\mu}\bar{\theta})_{\alpha }
\partial _{\mu }~,\qquad\bar{D}_{\dot{\alpha}}\,=\,i\partial _{\dot{\alpha}}~,
\end{eqnarray}%%
If we impose the chirality condition
\begin{eqnarray}\label{ecs2}
\bar{D}_{\dot{\alpha}}\Phi(x,\theta ,\bar{\theta})\!\!&=\!\!&0
\end{eqnarray}%
one gets the realization on Grassmann-holomorphic variables $\theta^{\alpha }$
($\alpha=1,2$) which enters into Euclidean $N=1$ holomorphic chiral superfields (see e.g.
\cite{LN84}). We obtain the following chiral differential realization of Euclidean
charges and supercharges by adjusting to Euclidean case so-called Mandelstam realization
\cite{Mandel,LN}
\begin{eqnarray}\label{ecs3}
\begin{array}{rcccl}
P_{\mu}\!\!&=\!\!&-i\partial_{\mu}~,\qquad \;\;M_{\mu\nu }\!\!&=\!\!&i
\displaystyle(x_{\mu}\partial_{\nu}-x_{\nu}\partial_{\mu})+\frac{1}{2}
(\theta\sigma_{\mu\nu})^{\alpha}\partial_{\alpha}~, %%
\\[10pt] %%
Q_{\alpha}\!\!& =\!\!&i\partial_{\alpha}~,\qquad\qquad\bar{Q}_{\dot{\alpha}}\!\!&=\!\!&
2(\theta\sigma^{\mu})_{\dot{\alpha}}\partial_{\mu}~,%
\end{array}%
\end{eqnarray}%
Inserting the realization (\ref{ecs3}) into the formulae (\ref{i15}), and using the
formula (\ref{i6}) extended to supertwist deformations of chiral superfields
\begin{eqnarray}\label{ecs4}
\Phi (x,\theta )\star_{{\cal F^{}}}\Psi(x,\theta)\!\!&:=\!\!&m\Bigl({\cal F}_{{}}^{-1}
\triangleright\bigl(\Phi(x,\theta)\otimes\Psi(x,\theta)\bigr)\Bigr)!,
\end{eqnarray}%
(the action $\triangleright$ is described by  the realization (\ref{ecs3})) we obtain
the following set of deformed superspace relations:

\noindent{\large\textit{(1). Deformation of Euclidean space-time}}. We use the notation
\begin{eqnarray}\label{ecs5}
[x^{\mu}_{},\,x^{\nu}_{}]_{\star_{\mathcal{F}_{ws}}}\!\!&:=\!\!&
x^{\mu}_{}\star_{\mathcal{F}_{ws}}x^{\nu}_{}-x^{\nu}_{}\star_{\mathcal{F}_{ws}}
x^{\mu}_{}~.
\end{eqnarray}
One can show that the general structure of this deformed commutator has the form
\begin{eqnarray}\label{ecs6}
[x^{\mu}_{},\,x^{\nu}_{}]_{\star_{\mathcal{F}_{ws}}}\!\!&=\!\!&
[x^{\mu}_{},\,x^{\nu}_{}]_{\star_{F_{1w}}}+[x^{\mu}_{},\,x^{\nu}_{}]_{%
\star_{F_{0}}}+[x^{\mu}_{},\,x^{\nu}_{}]_{\star_{F_{s}}}
\end{eqnarray}
for $w=a,b,c,d$. This property follows from the relations $P_{\mu}P_\nu\rhd
x^{\lambda}_{}=0$, $(Q^{\alpha})^2=0$. The explicit calculations yield to following
results:
\begin{eqnarray} \label{ecs7}
[x^{\mu}_{},\,x^{\nu}_{}]_{\star_{F_{s}}}\!\!&=\!\!&0~, %%
\\[10pt]\label{ecs8} %%
[x^{\mu}_{},\,x^{\nu}_{}]_{\star_{F_{0}}}\!\!&=\!\!&\frac{2i}{\kappa^2}\bigl(\alpha_1
\delta_{3}^{[\mu}\delta_0^{\nu]}+\alpha_2\delta_2^{[\mu}\delta_1^{\nu]}\bigr)~,
\\[10pt]\label{ecs9}
[x^{\mu}_{},\,x^{\nu}_{}]_{\star_{F_{1a}}}\!\!&=\!\!&\frac{2i\beta}{\kappa}
\bigl(\delta^{[\mu}_2\delta_0^{\nu]}x^{1}+\delta_0^{[\mu}\delta_1^{\nu]}x^{2}\bigr)~,
\\[10pt]\label{ecs10}
[x^{\mu}_{},\,x^{\nu}_{}]_{\star_{F_{1b}}}\!\!&=\!\!&\frac{2i\beta}{\kappa}
\bigl(\delta^{[\mu}_2\delta_3^{\nu]}x^{1}+\delta_3^{[\mu}\delta_1^{\nu]}x^{2}\bigr)~,
\\[10pt]\label{ecs11}
[x^{\mu}_{},\,x^{\nu}_{}]_{\star_{F_{1c}}}\!\!&=\!\!&
[x^{\mu}_{},\,x^{\nu}_{}]_{\star_{F_{1a}}}+[x^{\mu}_{},\,x^{\nu}_{}]_{ \star_{F_{1b}}}~,
\\[10pt]\label{ecs12}
[x^{\mu}_{},\,x^{\nu}_{}]_{\star_{F_{1d}}}\!\!&=\!\!&\frac{2i\beta}{\kappa}
\bigl(\delta_1^{[\mu} \delta_3^{\nu]}x^{0}+\delta_1^{[\mu}\delta_0^{\nu]}x^{3}\bigr).
\end{eqnarray}
It should be pointed out that only if we use in (5.4) the Mandelstam chiral realization
(\ref{ecs3}) the commutators of spacetime coordinates  %% (\ref{ecs7}--\ref{ecs12})
do not depend on the parameter $\eta$ and second Grassmann spinor
$\bar\theta^{\dot\alpha}$. If we consider the usual chiral realization which is obtained
by introducing the standard chiral fields as follows: $\tilde{\Phi}
(x,\theta,\bar{\theta})=\Phi(x-\imath\bar{\theta}\sigma\theta,\theta)$, the space-time
commutators will depend also on bilinear product $\bar{\theta} \sigma\theta$ of Grassmann
variables $\theta^{\alpha},\, \bar{\theta}^{\dot{\alpha}}$. Further we observe that

\textit{(i)} (\ref{ecs7}) describes a particular choices of the canonical deformation;

\textit{(ii)} the relations (\ref{ecs9})--(\ref{ecs12}) provide particular examples of
the Lie-algebraic deformations of spacetime coordinates;

\textit{(iii)} from the relation (\ref{ecs10}) follows that after deformation the time
coordinate remains commutative but other relations (\ref{ecs9}) and (\ref{ecs12})
describe the examples with quantum noncommutative time coordinate.

\noindent{\large{\textit{(2). Deformation of Grassmann sector}}}. The $\star$-product of
the chiral Grassmann variables is the same for all four deformations described by the
twists $\mathcal{F}_{ws}$ ($w=a,b,c,d$), and we obtain the following result
($\{\theta^{\alpha}_{},\,\theta^{\beta}_{}\}_{\star_{\mathcal{F}_{ws}}}=
\theta^{\alpha}_{}\star_{\mathcal{F}_{ws}}\theta^{\beta}_{}+
\theta^{\beta}_{}\star_{\mathcal{F}_{ws}}\theta^{\alpha}_{}$):
\begin{eqnarray}\label{ecs13}
\{\theta^{\alpha}_{},\,\theta^{\beta}_{}\}_{\star_{\mathcal{F}_{ws}}}\!\!&=\!\!&
\{\theta^{\alpha}_{},\,\theta^{\beta}_{}\}_{\star_{\mathcal{F}_{s}}}\,=\,-
\frac{2\eta}{\kappa}\delta^{(\alpha}_1\delta^{\beta)}_2~.
\end{eqnarray}
The relation (\ref{ecs13}) leads in the sector of Grassmann variables to the choice of
$\star$-product postulated by Seiberg \cite{Seib} which was introduced however as an
assumption without any link with the notation of quantum-deformed space-time symmetries.

\noindent{\large\textit{(3). Deformation of mixed spacetime-Grassmann sector}}. The
$\star$-product of Grassmann and spacetime coordinates depends on choice of the twist
$\mathcal{F}_{ws}$ ($w=a,b,c,d$) in the following way:
\begin{eqnarray}\label{ecs14}
[x_{}^{\mu },\,\theta_{}^{\alpha}]_{\star_{\mathcal{F}_{ws}}}\!\!&=\!\!&
[x_{}^{\mu},\,\theta_{}^{\alpha}]_{\star_{F_{1w}}}\qquad (w=a,b,c,d)~,
\end{eqnarray}%
where
\begin{eqnarray}\label{ecs15}
\begin{array}{rcccl}
[x_{}^{\mu},\,\theta_{}^{\alpha}]_{\star_{F_{1a}}}\!\!&=\!\!&
\displaystyle\frac{\beta}{\kappa}\delta_{0}^{\mu}(\theta\sigma_{12})^{\alpha},\qquad
\qquad\;[x_{}^{\mu},\,\theta_{}^{\alpha}]_{\star_{F_{1b}}}\!\!& =\!\!&
\displaystyle\frac{\beta}{\kappa}\delta_{3}^{\mu}(\theta\sigma_{12})^{\alpha}, %%
\\[10pt] %%
[x_{}^{\mu},\,\theta_{}^{\alpha}]_{\star_{F_{1c}}}\!\!& =\!\!&\displaystyle\frac{\beta}
{\kappa}\bigr(\delta_{0}^{\mu}+\delta_{3}^{\mu}\bigr)(\theta\sigma_{12})^{\alpha},\quad
\lbrack x_{{}}^{\mu},\,\theta_{{}}^{\alpha}]_{\star_{F_{1d}}}\!\!& =\!\!&
\displaystyle\frac{\beta}{\kappa}\delta_{1}^{\mu}(\theta\sigma_{03})^{\alpha},%
\end{array}%
\end{eqnarray}%
We see from the relations (\ref{ecs13})--(\ref{ecs15}) that only the chiral superspace
coordinates $\theta^{\alpha }$ are deformed, and the antichiral described by
$\bar{\theta}^{\dot{\alpha}}$-sector remains unchanged i.e. one obtained $N={1\over 2}$
SUSY deformation see \cite{Seib}--\cite{Buch1}. Such form of deformation is consistent
only in Euclidean framework, where the left-chiral and right-chiral coordinates can be
deformed independently.

We recall that first $N={1\over 2}$ deformation of Euclidean supersymmetries was
described by Seiberg in \cite{Seib} without any use of quantum supersymmetries. The
primary deformation in \cite{Seib} (and followed by other authors
\cite{Ivan}--\cite{Buch1}) is introduced by ansatz modifying anticommutator of half of
Grassmann variables
$$
\{\bar{\theta}_{\dot{\alpha}},\bar{\theta}_{\dot{\beta}}\}\,=\,0\Longrightarrow
\{\bar{\theta}_{\dot{\alpha}},\bar{\theta}_{\dot{\beta}}\}\,=\,\eta
C_{\dot{\alpha}\dot{\beta}}
$$
with other anti-commutators ($\{\theta_{\alpha},\theta_{\beta}\}=
\{\bar{\theta}_{\dot{\alpha}},\theta_{\beta}\}=0$) left unchanged. Inserting such
deformed Grassmann variables into the superspace realizations (\ref{ecs3}) we obtain the
following deformed anticommutators of antichiral supercharges
$$
\{\bar{Q}_{\dot{\alpha}},\bar{Q}_{\dot{\beta}}\}= \eta\delta_{\dot\alpha\dot\beta}\square
\equiv\eta\delta_{\dot\alpha\dot\beta} \delta^{\mu\nu}\partial_\mu\partial_\nu
$$
with unchanged anticommutators $\{Q_{\alpha}, Q_{\beta} \}$ and $\{\bar Q_{\dot\alpha},
Q_{\beta} \}$. In such a framework we obtain the $N={1\over 2}$ deformation of superspace
superalgebra which breaks the standard $D=4$ Euclidean supersymmetry.

For getting the modified Grassmann variables as in formula (\ref{ecs13}) it is sufficient
to consider the simplest canonical supertwist, described by the supersymmetric
$r$-matrices ${\cal N}=19-21$ in Table 2. Indeed such twist were considered for such
purpose in \cite{Zup} and \cite{GW1112} and they lead to the noncommutativity of
spacetime described by a constant matrix $\vartheta^{\mu\nu}$. The novelty of our results
here is the use of supersymmetric $r$-matrices with ${\cal N}=13-16$, which leads to
Lie-algebraic deformations of the spacetime sector, i.e. non-vanishing parameters
$\vartheta^{\mu\nu}_\rho$ in formula (\ref{i1}).

\section{Conclusions}

In this presentation we employed the formulae for the  $D=4$ Poincar\'{e} classical
$r$-matrices, obtained by Zakrzewski \cite{Z97}, and considered the corresponding $D=4$
Euclidean classical $r$-matrices. It appears that in Euclidean case due to the reality
condition (\ref{cr6}) only some of the  $r$-matrices from Table 1 can be used as the real
Euclidean classical $r$-matrices. Subsequently we did show also how to supersymmetrize
the $D=4$ Poincar\'{e} and Euclidean $r$-matrices (see Table 2) and considered the
restrictions imposed by the Poincar\'{e} and Euclidian reality conditions. Further for
four chosen supersymmetric $r$-matrices we constructed corresponding supertwists and
described respective quantum deformations of Euclidean superspace.

We made an important step in the task of providing the complete classification of
Hopf-algebraic quantum deformations of $D=4$ relativistic supersymmetries and their
Euclidean counterparts. We recall however that some of the considered supertwists violate
the reality condition in Minkowski superspace, but they are consistent with the reality
structure of Euclidean superalgebra, which is  described by the covariance under the
pseudoconjugation. We recall also that the known example of $D=4$ quantum deformations of
supersymmetries with Lie-algebraic deformed spacetime, the $\kappa$-deformation
\cite{LNS,KLMS} does not belong to the considered class of triangular quantum
superalgebras, because they can not be generated by Drinfeld twist\footnote{The exact
description of $\kappa-$deformed symmetries by twisting is only possible if we go beyond
the framework of standard quantum groups \cite{YZ}.}. It should be added that besides the
structure of twisted Poincar\'{e} and Euclidean superalgebras one can also consider
deformations of a dual Poincar\'{e} \cite{KLMS} as well as $D=4$ Euclidean \cite{GW1112}
supergroups.

In this presentation we considered explicitly  new examples of Hopf-algebraic framework
of twist-deformed $D=4$ supersymmetries and described corresponding new quantum
deformations of superspace. We stress that the considered class of deformed superspaces
is new with the Minkowski quantum algebra satisfying Lie-algebraic relations (see
(\ref{ecs8})-(\ref{ecs12}). The next step in our studies will be the introduction of
superfields on twist-deformed superspaces and the construction of new deformed
field-theoretical SUSY models with new quantum supersymmetries.

\subsection*{Acknowledgments} %%
The authors would like to thank Joseph Buchbinder for valuable remarks. The paper has
been supported by the the Polish National Science Center project 2011/01/B/ST2/03354
(A.B. and J.L.) and the grants RFBR-11-01-00980-a, NRU HSE-12-09-0064,
RFBR-09-01-93106-NCNIL-a (V.N.T.).

\end{document}